# *Decoding the International System*

Ingo Piepers

ingopiepers@gmail.com

**Abstract**

The International System develops according to a clear logic: By means of systemic wars organizational innovations are periodically introduced, contributing to a process of social expansion and integration, and to wealth creation. A finite-time singularity accompanied by four accelerating log-periodic cycles can be identified during the time frame 1495-1945.

This paper is based on two previous studies:

- *War: Origins and Effects*, Ingo Piepers, http://arxiv.org/abs/1409.6163
- *Regularities in the dynamics and development of the International System*, Ingo Piepers, http://arxiv.org/abs/1410.6477

Social scientists and historians have frequently attempted to determine whether regularities can be identified in the war dynamics of the International System. Until now, such efforts have not been successful.
However, with the help of complex systems theory, network science, and recently acquired insights in complex systems, it is now possible to identify various – and quite remarkable - regularities (4)(5)(10)(11)(12). These regularities provide us with information on ways to better manage the International System while more effectively preventing war.
These regularities necessitate the identification of a finite-time singularity that is accompanied by log-periodic oscillations in the long-term war dynamics of the International System.
To identify this singularity and other regularities, I apply a dataset prepared by Levy (7). In his study, Levy focuses on the war dynamics of 'Great Powers' (*see: Supplementary materials*). Great Powers can be accurately defined and identified based on their military capabilities and by their ability to project these capabilities. "A relatively high proportion of Great Power alliance commitments and war behavior is with each other, and Great Powers tend to perceive international relations as largely dependent upon and revolving around their own interrelationships. The general level of interactions among the Great Powers tends to be higher than for other states, whose interests are narrower and who interact primarily in more restricted regional settings. The Great Powers constitute an interdependent system of power and security relations." (7, p8-9).
The finite-time singularity of the International System is composed of four systemic wars. A number of characteristics distinguish systemic wars from non-systemic wars. Historians have also defined the four systemic wars as 'different' wars: wars with a significant and enduring impact on the functioning of the International System. These four systemic wars are the Thirty Years' War (1618-1648), the French Revolutionary and Napoleonic Wars (1792-1815), the First World War (1914-1918), and the Second World War (1939-1945).

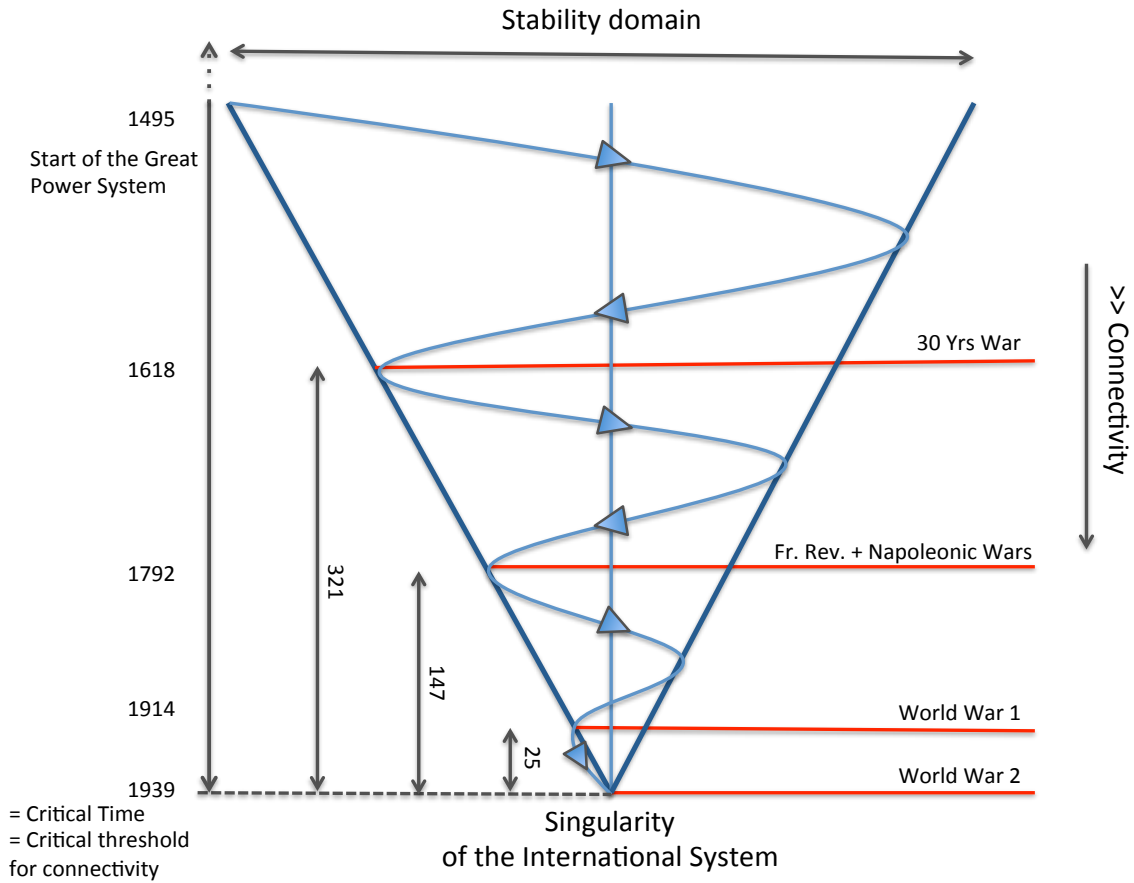

*Figure 1: This figure is a schematic representation of the finite-time singularity accompanied by log-periodic oscillations that shaped war dynamics and the development of the International System from 1495 - 1945.*

Systemic wars differ from non-systemic wars in several ways. Systemic wars typically affect the entire system. All Great Powers that form the Great Power System at the time of the systemic war actively participate in such wars.

Systemic wars are exceptionally high in intensity. Intensity is defined as "battle deaths per million European population" (7, 1983). A review of the intensities of successive systemic wars also reveals a clear pattern that suggests coherence between systemic wars: the intensity of successive systemic wars increases exponentially over time (*see Supplementary materials*).

This singularity, its dynamics, and the development of certain features of the International System suggest that the outcomes of systemic wars lead to the introduction of new and 'innovative' organizing principles for International System management. Successive organizing principles introduced from 1495 - 1945 progressively improved communication and coordination between the Great Powers.

Through the four successive systemic wars, the sovereignty and balance of power principles (Thirty Years' War) and 'platforms' for improving communication and coordination, including the Concert of Europe (French Revolutionary and Napoleonic Wars) and the League of Nations (First World War), were introduced. The Second World War represented a critical transition that resulted in a more fundamental reorganization of the International System, as I will explain below.

A final remarkable characteristic of systemic wars is their regularity in time. The timing of successive systemic wars, or oscillations, that form the singularity can be described through the following mathematical equation: *Life span (t) = $19.6e^{0.936 t}$*, with $R^2$ = 0.99 (*see: Supplementary materials*).

I assume that connectivity growth has served as the driver of the International System singularity (1)(10). Connectivity growth is stimulated by exponential population growth (6) and by the multiplication and growth of numerous social systems that are related to various human needs and social activities. Humans organize in groups and organizations to fulfill (basic) needs (2), differentiate their activities, achieve economies of scale, and generate more wealth; differentiation and economies of scale both increase wealth and ultimately improve the chances of survival.

The increased connectivity of the International System over time has increased the pace of life while exponentially shortening the life span of successive cycles. Cycles are defined as quasi-stable periods between successive systemic wars. Cycles provide opportunities for growth that are needed to sustain a constantly growing population while also generating opportunities for wealth creation.

Connectivity growth appears to encourage two competing 'forces' in the International System. On one hand, connectivity growth drives and enables cooperation and wealth creation. On the other hand, connectivity growth feeds discontent and causes imbalance, occasionally resulting in non-systemic wars.

The singularity dynamics and systemic wars that these competing forces produce are instrumental to the introduction of organizational innovation into the anarchistic International System. These innovations, which are being introduced at an accelerating rate, are needed to sustain growth and wealth creation. Singularity dynamics are defined as the inability of the anarchistic International System to reorganize itself through means other than systemic wars.

Thus, connectivity growth in an anarchistic system has two major consequences: large-scale destruction on one hand, and the introduction of new forms of cooperation necessary for growth and wealth creation on the other hand.

At the critical period of the singularity, 1939, International System connectivity reached a critical threshold. Before this threshold was reached, the International System could periodically 'reset its parameters' within the context of the anarchistic International System while producing a new quasi-stable cycle that allowed for further growth and wealth creation. However, at this critical time, International System connectivity became incompatible with the anarchistic nature of the International System, and a critical transition became both necessary and unavoidable.

The critical transition resulting from the fourth systemic war, the Second World War, had two effects: the transformation of the anarchistic European System into a cooperative security community (3), and the 'globalization' of the International System.

After the Second World War, Europe was primed for a new level of cooperation, and a fundamentally new organizational structure and 'management' (8) approach could be introduced.

In fact, the singularity dynamic was - in the period 1495 - 1945 - instrumental in a process of social expansion and integration that already started millennia. The occurrence of social expansion and integration during this period (accelerated cycles of innovation that improved cooperation as introduced by a series of systemic wars) can be attributed to the anarchistic nature of the International System.

The singularity generates a framework for identifying four 'cycles' of war dynamics from 1495-1945. I define other wars that occurred during these cycles as 'non-systemic wars.' It is now possible to further examine the features of these cycles.

| Evolution of successive cycle stability and resilience | | | | | |
|---|---|---|---|---|---|
| | | Stability | | Resilience | |
| Cycle | Period | War frequency | Great Power status dynamics | Number of Great Power wars | Life span (years) |
| 1 | 1495 – 1618 | 0.37 | 8 | 45 | 123 |
| 2 | 1648 – 1792 | 0.24 | 5 | 34 | 144 |
| 3 | 1815 – 1914 | 0.18 | 3 | 18 | 99 |
| 4 | 1918 – 1939 | 0.05 | 0 | 1 | 21 |

*Table 1: This table presents data on successive cycles to support an analysis of International System war dynamics and development (see: Supplementary materials).*

The analysis shows a linear decrease in the frequency of war in successive cycles. I use war frequency development as a measure of the evolution of International System stability over time. Stability is defined as the ability of the International System to sustain a state of peace, that is, the absence of Great Power wars. A linear decrease in war frequency reflects a linear increase in International System stability. It is likely not coincidental that Great Power status dynamics during successive cycles also show a linear decrease: stability and status dynamics are closely related. I define International System *status dynamics* as the number of states that acquire or lose their Great Power status during successive cycles.

Based on Levy's dataset, it is possible to conclude that status dynamics decreased linearly over time (7, 1983, p47). During the first four cycles, eight, five, three, and zero status changes occurred, respectively (status changes that occurred during systemic wars are excluded).

Two of the three status changes that occurred during the third cycle involved the United States (1898) and Japan (1905). The evolution of this indicator overtime not only reveals an increase in European system stability but also the heightened impact of non-European states on International System dynamics.

The absolute number of non-systemic wars that have occurred during successive cycles can be defined as a measure of the development of International System resilience over time. I define resilience as the capacity for the system to sustain itself within a particular stability domain or within a certain 'cycle.' The absolute number of wars decreases linearly, reflecting a linear decrease in International System resilience.

It appears that a linear increase in International System stability occurred in conjunction with a linear decrease in system resilience. Thus, a more stable International System does not necessary reflect a more secure system.

The development of International System stability and resilience may also be related to system connectivity growth.

I assume that increased connectivity generates greater stability: Great Powers become more 'tightly' linked, decreasing the frequency of war.

As shown above, a linear increase in stability corresponds to a linear decline in resilience. This relationship suggests that a more connected and stable International System is less likely to tolerate disturbances before the system must reorganize.

As International System stability increased linearly, the intensity of successive systemic wars grew exponentially. Rather, a more connected and consequently more stable system necessitated more 'energy' inputs to rebalance and re-establish the system within a new viable stability domain (a new quasi-stable cycle in this context).

The life span of the first three systemic wars also shows an exponential decrease in conjunction with an exponential shortening of the life span of cycles. Connectivity growth appears to exponentially increase the pace of life not only during quasi-stable cycles but also during successive systemic wars that 'accompany' them.

These findings have profound consequences. They show, for example, that to understand social dynamics and historical events, it does not suffice to merely examine the social dynamics of the International System network. Rather, an analysis of the dynamics of the underlying network and of the interplay between both dynamic levels is necessary.

The social dynamics of the International System network constitute a series of unique events, whereas the dynamics of the underlying network show remarkable regularities. These regularities shape the social dynamics and the development of the International System.

It is now necessary to determine whether the current International System of the post-1939 singularity exhibits similar dynamics to its predecessor. This perspective appears to be a plausible and wise assumption, as the current International System is also an anarchistic system in which connectivity continues to grow, likely exponentially. Thus, the destructive system of connectivity growth in an anarchistic system remains intact.

In the current International System, we still lack a 'management' structure for reorganizing the International System, introducing new organizing principles, or for allowing growth and wealth creation without resorting to systemic war. We should now focus our efforts on developing and peacefully implementing a new governance structure for the International System to avoid large-scale and likely irreparable damage.

The highly deterministic nature of the International System and the regularities it produced (at least in the past) also allow for the identification of early-warning signals and for the establishment of a monitoring system (9). History and highly destructive capabilities that are now enabled through the International System should serve as sufficient incentives to take action.

# *Supplementary materials*

## *Decoding the International System*

Version: March 1st 2015

Ingo Piepers

ingopiepers@gmail.com

### 1. **Levy's dataset**

The tables below show Levy's dataset, including an additional unit of measurement for the size of Great Power Wars: 'Fraction'.

Levy has defined (7, p81, p92) the following units of measurement: *Duration*: years; *Extent*: number of Powers; *Magnitude*: nation-years; *Severity*: the number of battle-connected deaths of military personnel; *Intensity*: battle deaths per million European population; and *Concentration*: battle fatalities per nation year.

*Fraction* is the unit of measurement I have introduced for the size of Great Power wars. Size of a particular Great Power war is measured relative to the size of the Great Power System at that particular moment in time. This measure is calculated by dividing the number of Great Powers involved in a war, by the total number of Great Powers that exist at that moment in time in the International System.

The wars that are marked red constitute systemic wars: Numbers 46-49 are the Thirty Years' War, 84-85 the French Revolutionary and Napoleonic Wars, 107 the First World War, and 113 the Second World War.

| Nr. Levy | Start | End | Duration | Number GP | Extent | Fraction | Magnitude | Concentration | Intensity | Severity |
|---|---|---|---|---|---|---|---|---|---|---|
| 1 | 1495 | 1497 | 2,0 | 5 | 3 | 0,60 | 1,20 | 1333 | 119 | 8000 |
| 2 | 1497 | 1498 | 1,0 | 5 | 1 | 0,20 | 0,20 | 3000 | 45 | 3000 |
| 3 | 1499 | 1503 | 4,0 | 5 | 1 | 0,20 | 0,80 | 1000 | 60 | 4000 |
| 4 | 1499 | 1500 | 1,0 | 5 | 1 | 0,20 | 0,20 | 2000 | 29 | 2000 |
| 5 | 1501 | 1504 | 3,0 | 5 | 2 | 0,40 | 1,20 | 3600 | 269 | 18000 |
| 6 | 1508 | 1509 | 1,0 | 5 | 3 | 0,60 | 0,60 | 3333 | 145 | 10000 |
| 7 | 1511 | 1514 | 3,0 | 5 | 4 | 0,80 | 2,40 | 1500 | 261 | 18000 |
| 8 | 1512 | 1519 | 7,0 | 5 | 2 | 0,40 | 2,80 | 1714 | 343 | 24000 |
| 9 | 1513 | 1515 | 2,0 | 5 | 1 | 0,20 | 0,40 | 2000 | 57 | 4000 |
| 10 | 1515 | 1515 | 0,5 | 5 | 3 | 0,60 | 0,30 | 2000 | 43 | 3000 |
| 11 | 1521 | 1526 | 5,0 | 4 | 3 | 0,75 | 3,75 | 2000 | 420 | 30000 |
| 12 | 1521 | 1531 | 10,0 | 4 | 2 | 0,50 | 5,00 | 3400 | 958 | 68000 |
| 13 | 1522 | 1523 | 1,0 | 4 | 1 | 0,25 | 0,25 | 3000 | 41 | 3000 |
| 14 | 1526 | 1529 | 3,0 | 4 | 3 | 0,75 | 2,25 | 2250 | 249 | 18000 |
| 15 | 1532 | 1535 | 3,0 | 4 | 2 | 0,50 | 1,50 | 4667 | 384 | 28000 |
| 16 | 1532 | 1534 | 2,0 | 4 | 1 | 0,25 | 0,50 | 2000 | 55 | 4000 |
| 17 | 1536 | 1538 | 2,0 | 4 | 2 | 0,50 | 1,00 | 8000 | 438 | 32000 |
| 18 | 1537 | 1547 | 10,0 | 4 | 2 | 0,50 | 5,00 | 4850 | 1329 | 97000 |
| 19 | 1542 | 1550 | 8,0 | 4 | 1 | 0,25 | 2,00 | 1625 | 176 | 13000 |
| 20 | 1542 | 1544 | 2,0 | 4 | 2 | 0,50 | 1,00 | 11750 | 629 | 47000 |
| 21 | 1544 | 1546 | 2,0 | 4 | 2 | 0,50 | 1,00 | 2000 | 107 | 8000 |
| 22 | 1549 | 1550 | 1,0 | 4 | 2 | 0,50 | 0,50 | 3000 | 79 | 6000 |
| 23 | 1551 | 1556 | 5,0 | 4 | 2 | 0,50 | 2,50 | 4400 | 578 | 44000 |
| 24 | 1552 | 1556 | 4,0 | 4 | 2 | 0,50 | 2,00 | 6375 | 668 | 51000 |
| 25 | 1556 | 1562 | 6,0 | 5 | 2 | 0,40 | 2,40 | 4333 | 676 | 52000 |
| 26 | 1556 | 1559 | 3,0 | 5 | 3 | 0,60 | 1,80 | 3000 | 316 | 24000 |
| 27 | 1559 | 1560 | 1,0 | 5 | 2 | 0,40 | 0,40 | 4000 | 78 | 6000 |
| 28 | 1559 | 1564 | 5,0 | 5 | 2 | 0,40 | 2,00 | 2400 | 310 | 24000 |
| 29 | 1562 | 1564 | 2,0 | 5 | 2 | 0,40 | 0,80 | 1500 | 77 | 6000 |
| 30 | 1565 | 1568 | 3,0 | 5 | 2 | 0,40 | 1,20 | 4000 | 306 | 24000 |
| 31 | 1569 | 1580 | 11,0 | 5 | 2 | 0,40 | 4,40 | 2182 | 608 | 48000 |
| 32 | 1576 | 1583 | 7,0 | 5 | 2 | 0,40 | 2,80 | 3429 | 600 | 48000 |
| 33 | 1579 | 1581 | 2,0 | 5 | 1 | 0,20 | 0,40 | 2000 | 50 | 4000 |
| 34 | 1583 | 1590 | 7,0 | 5 | 1 | 0,20 | 1,40 | 2429 | 210 | 17000 |
| 35 | 1585 | 1604 | 19,0 | 5 | 2 | 0,40 | 7,60 | 1263 | 588 | 48000 |
| 36 | 1587 | 1588 | 1,0 | 5 | 1 | 0,20 | 0,20 | 4000 | 49 | 4000 |
| 37 | 1589 | 1598 | 9,0 | 5 | 2 | 0,40 | 3,60 | 889 | 195 | 16000 |
| 38 | 1593 | 1606 | 13,0 | 5 | 2 | 0,40 | 5,20 | 3462 | 1086 | 90000 |
| 39 | 1600 | 1601 | 1,0 | 5 | 1 | 0,20 | 0,20 | 2000 | 24 | 2000 |
| 40 | 1610 | 1614 | 4,0 | 6 | 2 | 0,33 | 1,33 | 1875 | 175 | 15000 |
| 41 | 1615 | 1618 | 3,0 | 6 | 1 | 0,17 | 0,50 | 2000 | 70 | 6000 |
| 42 | 1615 | 1617 | 2,0 | 6 | 1 | 0,17 | 0,33 | 1000 | 23 | 2000 |
| 43 | 1617 | 1621 | 4,0 | 7 | 1 | 0,14 | 0,57 | 1250 | 58 | 5000 |
| 44 | 1618 | 1619 | 1,0 | 7 | 2 | 0,29 | 0,29 | 3000 | 69 | 6000 |
| 45 | 1618 | 1621 | 3,0 | 7 | 1 | 0,14 | 0,43 | 5000 | 173 | 15000 |
| 46 | 1618 | 1625 | 7,0 | 7 | 4 | 0,57 | 4,00 | 20267 | 3535 | 304000 |
| 47 | 1625 | 1630 | 5,0 | 7 | 6 | 0,86 | 4,29 | 11615 | 3432 | 302000 |
| 48 | 1630 | 1635 | 5,0 | 7 | 4 | 0,57 | 2,86 | 15700 | 3568 | 214000 |
| 49 | 1635 | 1648 | 13,0 | 7 | 5 | 0,71 | 9,29 | 17708 | 12933 | 1151000 |
| 50 | 1642 | 1668 | 26,0 | 7 | 1 | 0,14 | 3,71 | 3077 | 882 | 80000 |
| 51 | 1645 | 1664 | 19,0 | 7 | 1 | 0,14 | 2,71 | 3790 | 791 | 72000 |
| 52 | 1648 | 1659 | 11,0 | 7 | 2 | 0,29 | 3,14 | 4909 | 1187 | 108000 |
| 53 | 1650 | 1651 | 1,0 | 7 | 1 | 0,14 | 0,14 | 2000 | 22 | 2000 |
| 54 | 1652 | 1655 | 3,0 | 7 | 2 | 0,29 | 0,86 | 4333 | 282 | 26000 |
| 55 | 1654 | 1660 | 6,0 | 7 | 3 | 0,43 | 2,57 | 1833 | 238 | 22000 |
| 56 | 1656 | 1659 | 3,0 | 7 | 2 | 0,29 | 0,86 | 2500 | 161 | 15000 |
| 57 | 1657 | 1661 | 4,0 | 7 | 1 | 0,14 | 0,57 | 1000 | 43 | 4000 |
| 58 | 1657 | 1664 | 7,0 | 7 | 3 | 0,43 | 3,00 | 8385 | 1170 | 109000 |
| 59 | 1665 | 1666 | 1,0 | 7 | 1 | 0,14 | 0,14 | 1000 | 11 | 2000 |
| 60 | 1665 | 1667 | 2,0 | 7 | 3 | 0,43 | 0,86 | 6167 | 392 | 37000 |

| Nr. Levy | Start | End | Duration | Number GP | Extent | Fraction | Magnitude | Concentration | Intensity | Severity |
|---|---|---|---|---|---|---|---|---|---|---|
| 61 | 1667 | 1668 | 1,0 | 7 | 2 | 0,29 | 0,29 | 2000 | 42 | 4000 |
| 62 | 1672 | 1678 | 6,0 | 7 | 6 | 0,86 | 5,14 | 10364 | 3580 | 342000 |
| 63 | 1672 | 1676 | 4,0 | 7 | 1 | 0,14 | 0,57 | 1250 | 52 | 5000 |
| 64 | 1677 | 1681 | 4,0 | 7 | 1 | 0,14 | 0,57 | 3000 | 125 | 12000 |
| 65 | 1682 | 1699 | 17,0 | 7 | 2 | 0,29 | 4,86 | 11294 | 3954 | 384000 |
| 66 | 1683 | 1684 | 1,0 | 7 | 2 | 0,29 | 0,29 | 2500 | 51 | 5000 |
| 67 | 1688 | 1697 | 9,0 | 7 | 5 | 0,71 | 6,43 | 15111 | 6939 | 680000 |
| 68 | 1700 | 1721 | 21,0 | 6 | 2 | 0,33 | 7,00 | 2370 | 640 | 64000 |
| 69 | 1701 | 1713 | 12,0 | 6 | 5 | 0,83 | 10,00 | 20850 | 12490 | 1251000 |
| 70 | 1716 | 1718 | 2,0 | 5 | 1 | 0,20 | 0,40 | 5000 | 98 | 10000 |
| 71 | 1718 | 1720 | 2,0 | 5 | 4 | 0,80 | 1,60 | 3125 | 245 | 25000 |
| 72 | 1726 | 1729 | 3,0 | 5 | 2 | 0,40 | 1,20 | 2500 | 144 | 15000 |
| 73 | 1733 | 1738 | 5,0 | 5 | 4 | 0,80 | 4,00 | 4400 | 836 | 88000 |
| 74 | 1736 | 1739 | 3,0 | 5 | 2 | 0,40 | 1,20 | 6333 | 359 | 38000 |
| 75 | 1739 | 1748 | 9,0 | 6 | 6 | 1,00 | 9,00 | 8159 | 3379 | 359000 |
| 76 | 1741 | 1743 | 2,0 | 6 | 1 | 0,17 | 0,33 | 5000 | 94 | 10000 |
| 77 | 1755 | 1763 | 8,0 | 6 | 6 | 1,00 | 8,00 | 26105 | 9118 | 992000 |
| 78 | 1768 | 1774 | 6,0 | 6 | 1 | 0,17 | 1,00 | 2333 | 127 | 14000 |
| 79 | 1768 | 1772 | 4,0 | 6 | 1 | 0,17 | 0,67 | 3500 | 149 | 14000 |
| 80 | 1778 | 1779 | 1,0 | 6 | 2 | 0,33 | 0,33 | 150 | 3 | 300 |
| 81 | 1778 | 1784 | 6,0 | 6 | 3 | 0,50 | 3,00 | 2267 | 304 | 34000 |
| 82 | 1787 | 1792 | 5,0 | 6 | 2 | 0,33 | 1,67 | 192000 | 1685 | 192000 |
| 83 | 1788 | 1790 | 2,0 | 6 | 1 | 0,17 | 0,33 | 1500 | 26 | 3000 |
| 84 | 1792 | 1802 | 10,0 | 6 | 6 | 1,00 | 10,00 | 13000 | 5816 | 663000 |
| 85 | 1803 | 1815 | 12,0 | 6 | 6 | 1,00 | 12,00 | 32224 | 16112 | 1869000 |
| 86 | 1806 | 1812 | 6,0 | 6 | 2 | 0,33 | 2,00 | 6429 | 388 | 45000 |
| 87 | 1808 | 1809 | 1,5 | 5 | 1 | 0,20 | 0,30 | 4000 | 51 | 6000 |
| 88 | 1812 | 1814 | 2,5 | 5 | 1 | 0,20 | 0,50 | 1600 | 34 | 4000 |
| 89 | 1815 | 1815 | 0,5 | 5 | 1 | 0,20 | 0,10 | 10000 | 17 | 2000 |
| 90 | 1823 | 1823 | 0,9 | 5 | 1 | 0,20 | 0,18 | 667 | 3 | 400 |
| 91 | 1827 | 1827 | 0,1 | 5 | 3 | 0,60 | 0,06 | 1800 | 2 | 180 |
| 92 | 1828 | 1829 | 1,0 | 5 | 1 | 0,20 | 0,20 | 35714 | 415 | 50000 |
| 93 | 1848 | 1849 | 1,0 | 5 | 1 | 0,20 | 0,20 | 5600 | 45 | 5600 |
| 94 | 1849 | 1849 | 1,2 | 5 | 1 | 0,20 | 0,24 | 2083 | 20 | 2500 |
| 95 | 1849 | 1849 | 0,2 | 5 | 2 | 0,40 | 0,08 | 1500 | 4 | 600 |
| 96 | 1853 | 1856 | 2,4 | 5 | 3 | 0,60 | 1,44 | 35000 | 1743 | 217000 |
| 97 | 1856 | 1857 | 0,4 | 5 | 1 | 0,20 | 0,08 | 1250 | 4 | 500 |
| 98 | 1859 | 1859 | 0,2 | 5 | 2 | 0,40 | 0,08 | 50000 | 159 | 20000 |
| 99 | 1862 | 1867 | 4,8 | 6 | 1 | 0,17 | 0,80 | 1667 | 64 | 8000 |
| 100 | 1864 | 1864 | 0,5 | 6 | 2 | 0,33 | 0,17 | 1500 | 12 | 1500 |
| 101 | 1866 | 1866 | 0,1 | 6 | 3 | 0,50 | 0,05 | 113333 | 270 | 34000 |
| 102 | 1870 | 1871 | 0,6 | 6 | 2 | 0,33 | 0,20 | 150000 | 1415 | 180000 |
| 103 | 1877 | 1878 | 0,7 | 6 | 1 | 0,17 | 0,12 | 171429 | 935 | 120000 |
| 104 | 1884 | 1885 | 1,0 | 6 | 1 | 0,17 | 0,17 | 2100 | 16 | 2100 |
| 105 | 1904 | 1905 | 1,6 | 7 | 1 | 0,14 | 0,23 | 28125 | 339 | 45000 |
| 106 | 1911 | 1912 | 1,1 | 8 | 1 | 0,13 | 0,14 | 5454 | 45 | 6000 |
| 107 | 1914 | 1918 | 4,3 | 8 | 8 | 1,00 | 4,30 | 258672 | 57616 | 7734300 |
| 108 | 1918 | 1921 | 3,0 | 7 | 5 | 0,71 | 2,14 | 385 | 37 | 5000 |
| 109 | 1931 | 1933 | 1,4 | 7 | 1 | 0,14 | 0,20 | 7143 | 73 | 10000 |
| 110 | 1935 | 1936 | 0,6 | 7 | 1 | 0,14 | 0,09 | 6667 | 29 | 4000 |
| 111 | 1937 | 1941 | 4,4 | 7 | 1 | 0,14 | 0,63 | 56819 | 1813 | 250000 |
| 112 | 1939 | 1939 | 0,4 | 7 | 2 | 0,29 | 0,11 | 22857 | 116 | 16000 |
| 113 | 1939 | 1945 | 6,0 | 7 | 7 | 1,00 | 6,00 | 462439 | 93665 | 12948300 |
| 114 | 1939 | 1940 | 0,3 | 7 | 1 | 0,14 | 0,04 | 166667 | 362 | 50000 |
| 115 | 1950 | 1953 | 3,1 | 5 | 4 | 0,80 | 2,48 | 84510 | 6821 | 954960 |
| 116 | 1956 | 1956 | 0,1 | 6 | 1 | 0,17 | 0,02 | 70000 | 50 | 7000 |
| 117 | 1956 | 1956 | 0,1 | 6 | 2 | 0,33 | 0,03 | 300 | 0 | 30 |
| 118 | 1962 | 1962 | 0,1 | 6 | 1 | 0,17 | 0,02 | 5000 | 1 | 500 |
| 119 | 1965 | 1973 | 8,0 | 6 | 1 | 0,17 | 1,33 | 7000 | 90 | 56000 |

*Table 1: Levy's dataset. Fraction (measure for size) added*

| 1 | War of the League of Venice* | 41 | Austro-Venetian War | 81 | War of the American Revolution* |
|---|---|---|---|---|---|
| 2 | Polish-Turkish War | 42 | Spanish-Savoian War | 82 | Ottoman War |
| 3 | Venitian-Turkish War | 43 | Spanish-Venetian War | 83 | Russo-Swedish War |
| 4 | First Milanese War | 44 | Spanish-Turkish War* | 84 | French Revolutionary Wars* |
| 5 | Neapolitan War* | 45 | Polish-Turkish War | 85 | Napoleonic Wars* |
| 6 | War of the Cambrian League | 46 | Thirty Year's War - Bohemian* | 86 | Russo-Turkish War |
| 7 | War of the Holy League* | 47 | Thirty Year's War - Danish* | 87 | Russo-Swedish War |
| 8 | Austro-Turkish War* | 48 | Thirty Year's War - Swedish* | 88 | War of 1812 |
| 9 | Scottish War | 49 | Thirty Year's War - Swedish-French* | 89 | Neapolitan War |
| 10 | Second Milanese War* | 50 | Spanish-Portuguese War | 90 | Franco-Spanish War |
| 11 | First War of Charles V* | 51 | Turkish-Venetian War | 91 | Navarino Bay |
| 12 | Ottoman War* | 52 | Franco-Spanish War* | 92 | Russo-Turkish War |
| 13 | Scottish War | 53 | Scottish War | 93 | Austro-Sardinian War |
| 14 | Second War of Charles V* | 54 | Anglo-Dutch Naval War* | 94 | First Schleswig-Holstein War |
| 15 | Ottoman War* | 55 | Great Northern War* | 95 | Roman Republic War |
| 16 | Scottish War | 56 | English-Sopanish War* | 96 | Crimean War* |
| 17 | Third War of Charles V* | 57 | Dutch-Portuguese War | 97 | Anglo-Perian War |
| 18 | Ottoman War* | 58 | Ottoman War* | 98 | War of Italian Unifification* |
| 19 | Scottish War | 59 | Sweden-Bremen War | 99 | Franco-Mexican War |
| 20 | Fourth War of Charles V* | 60 | Anglo-Dutch Naval War* | 100 | Second Schleswig-Holstein War |
| 21 | Siege of Boulogne* | 61 | Devolutionary War* | 101 | Austro-Prussian War* |
| 22 | Arundel's Rebellion* | 62 | Dutch War of Louis XIV* | 102 | Franco-Prusssian War* |
| 23 | Ottoman War* | 63 | Turkish-Polish War | 103 | Russo-Turkish War |
| 24 | Fifth War of Charles V* | 64 | Russo-Turkish War | 104 | Sino-French War |
| 25 | Austro-Turkish War* | 65 | Ottoman War* | 105 | Russo-Japanese War |
| 26 | Franco_Spanish War* | 66 | Franco-Spanish War* | 106 | Italo-Turkish War |
| 27 | Scottish War* | 67 | War of the League of Augusburg* | 107 | World War I* |
| 28 | Spanish-Turkish War* | 68 | Second Northern War* | 108 | Russian Civil War* |
| 29 | First Huguenot War* | 69 | War of the Spanish Succession* | 109 | Manchurian War |
| 30 | Austro-Turkish War* | 70 | Ottoman War | 110 | Italo-Ethiopian War |
| 31 | Spanish-Turkish War* | 71 | War of the Quadruple Alliance* | 111 | Sino-Japanese War |
| 32 | Austro-Turkish War* | 72 | British-Spanish War* | 112 | Russo-Japanese War* |
| 33 | Spanish-Potuguese War | 73 | War of the Polish Succession* | 113 | World War II* |
| 34 | Polish-Turkish War | 74 | Ottoman War | 114 | Russo-Finnish War |
| 35 | War of the Armada* | 75 | War of the Austrian Succession* | 115 | Korean War* |
| 36 | Austro-Polish War | 76 | Russo-Swedish War | 116 | Russo-Hungarian War |
| 37 | War of the Three Henries* | 77 | Seven Years' War* | 117 | Sinai War |
| 38 | Austro-Turkish War* | 78 | Russo-Turkish War | 118 | Sino-Indian War |
| 39 | Franco-Savoian War | 79 | Confederation of Bar | 119 | Vietnam War |
| 40 | Spanish-Turkish War* | 80 | War of the Bavarian Succession* |  |  |

*Table 2: Great Power Wars: Numbers and names*

## 2. Intensity of successive systemic wars

The intensities of successive systemic wars also shows remarkable regularities: the intensities of successive systemic wars increased exponentially. Levy defines *intensity* as "battle deaths per million European population" (1, 1983).
In this overview I did not include the fourth systemic war: the Second World War. The reason to exclude this particular systemic war is the fact that it constituted a critical transition, and cannot be 'compared' with the other systemic wars. However, the observation that the intensity grows exponentially still holds when the Second World War is also included in the analysis.
The mathematical expression for this regularity is: ***Intensity / year (t) = 5.68 e^(2.61t), with R2 = 1.00 (t = the number of the systemic war)*** and for the absolute intensity of successive systemic wars, where t also is the number of the systemic war:
***Absolute Intensity (t) = 8664.47 e^(0.62t), with R2 = 0.97.***

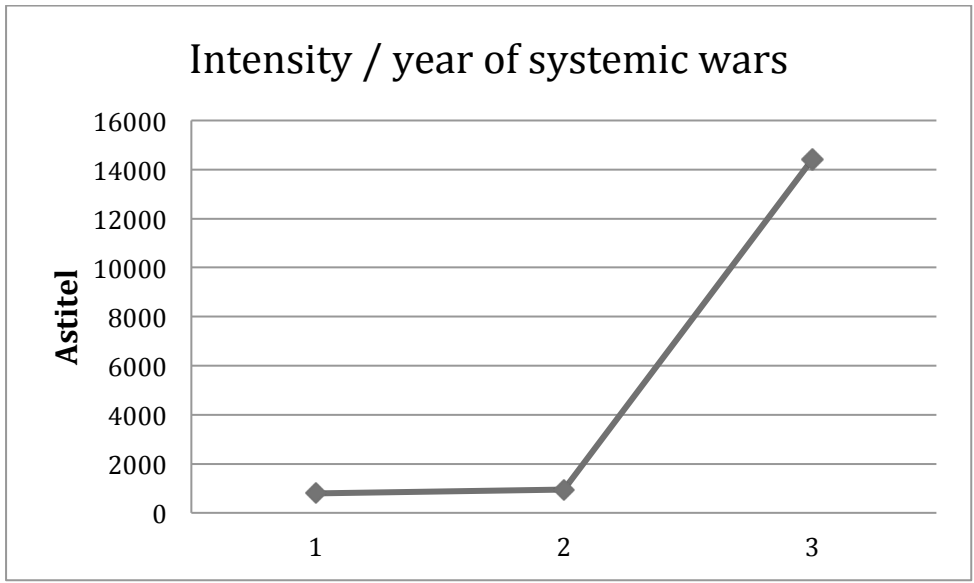

*Figure 1: Exponential increase of the ratio intensity / year of successive systemic wars.*

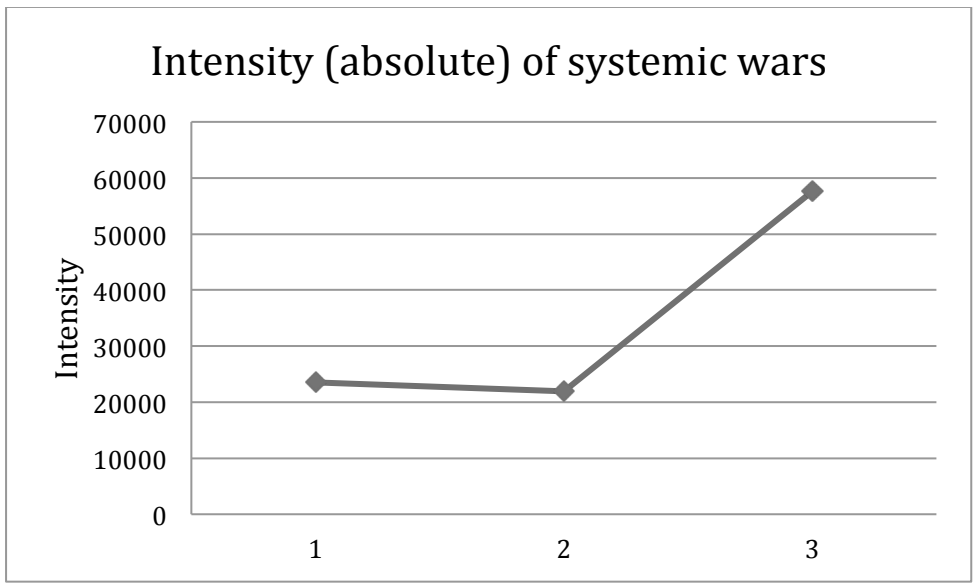

*Figure 2: Exponential increase of the absolute intensity of successive systemic wars.*

## 3. Singularity dynamics

| Evolution of the life spans of successive oscillations | | | |
|---|---|---|---|
| **Systemic War** | | **Time** | **t(c) – t** |
| Second World War | t(c) = t-critical | 1939 | 0 |
| First World War | t(1) | 1914 | 25 |
| French Revolutionary and Napoleonic Wars | t(2) | 1792 | 147 |
| Thirty Years' War | t(3) | 1618 | 321 |

*Table 3. Evolution of the life spans of successive oscillations*

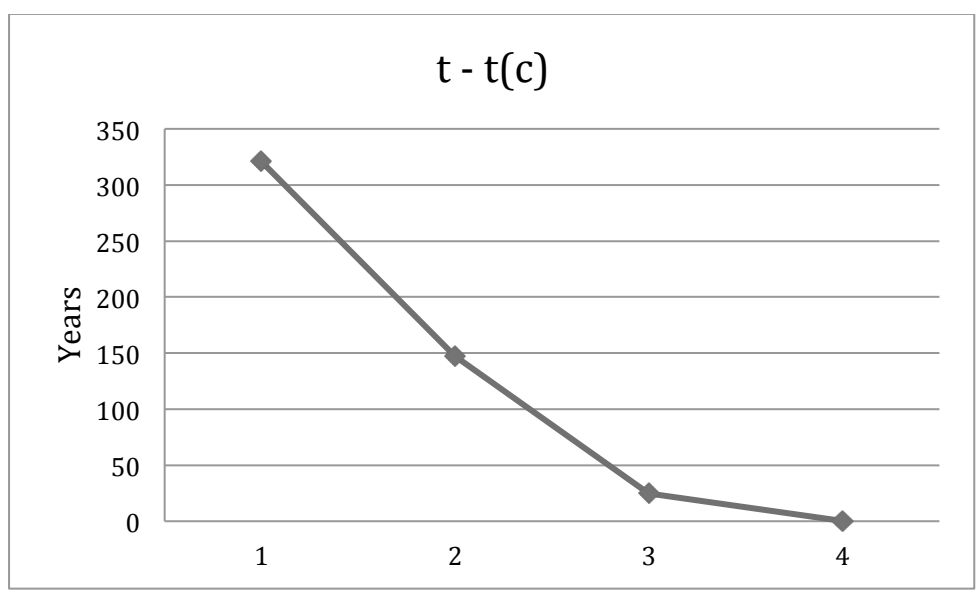

*Figure 3: Identification of a finite-time singularity accompanied by log-periodic oscillations in the war dynamics of the International System.*

## 4. Evolution of successive cycle stability and resilience

Table 1 in the article shows the stability and resilience of the four successive cycles; it only addresses non-systemic wars. Systemic wars are not included in this overview because they constitute a fundamentally different category. The number of wars in Successive International Systems and the war frequency, for example, evolve linearly. The war frequency of cycles is calculated by dividing the number of Great Power wars by life span of the cycle. Great Power wars outside the European continent, with only one European participant, are excluded from this overview. Thus, there are seven relevant wars involved (numbers 97, 99, 104, 105, 109, 110, and 111) in the 1856-1941 period. From another perspective, this result shows a different category of wars (wars outside Europe, European Great Powers in war with non-European states). These wars are indicative of the globalization of the International System but obscure the process of social expansion and integration in Europe.

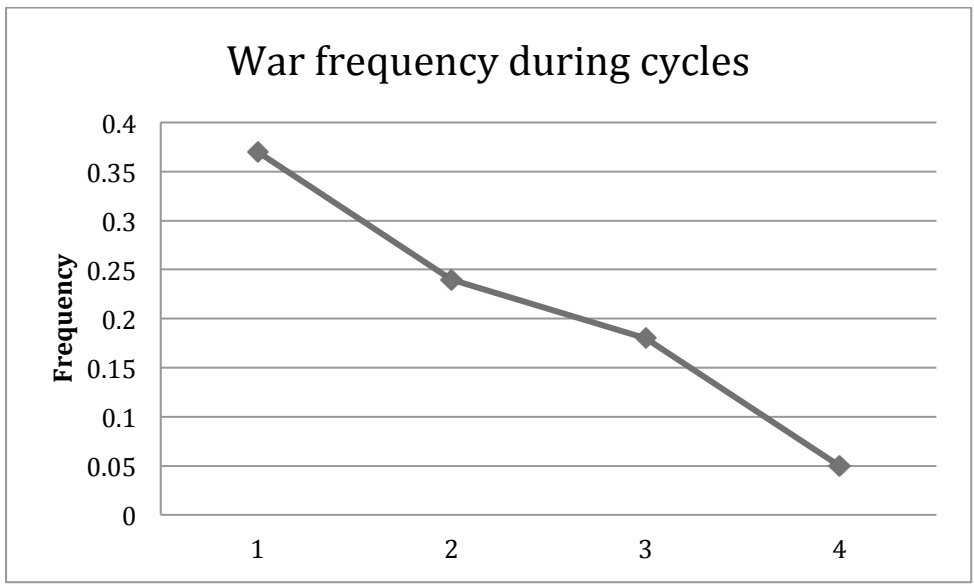

*Figure 4: This figure shows the linear decrease of the war frequency of the International System. This linear decrease in war frequency implies a linear increase in stability.*

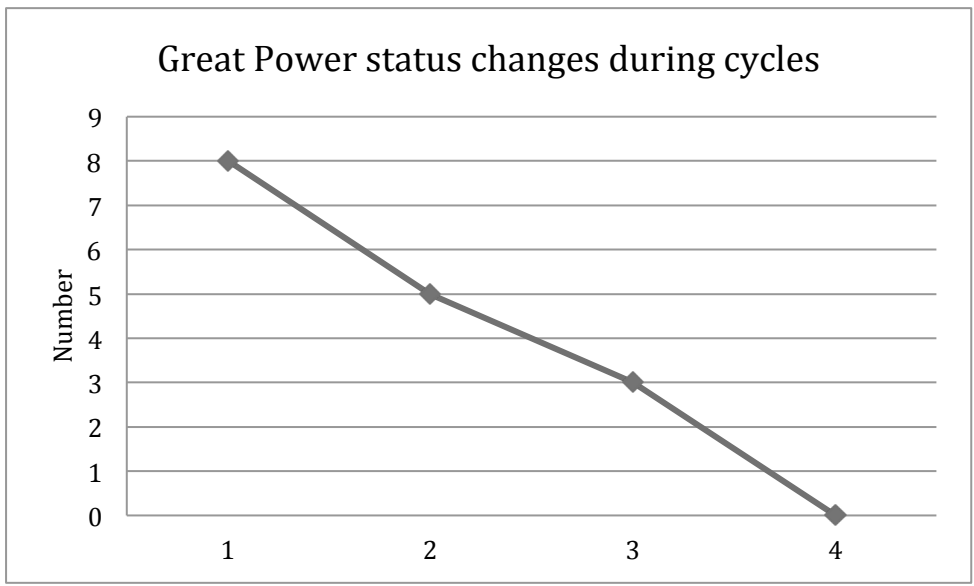

*Figure 5: This figure also shows the linear increase of the stability of the International System during four successive cycles.*

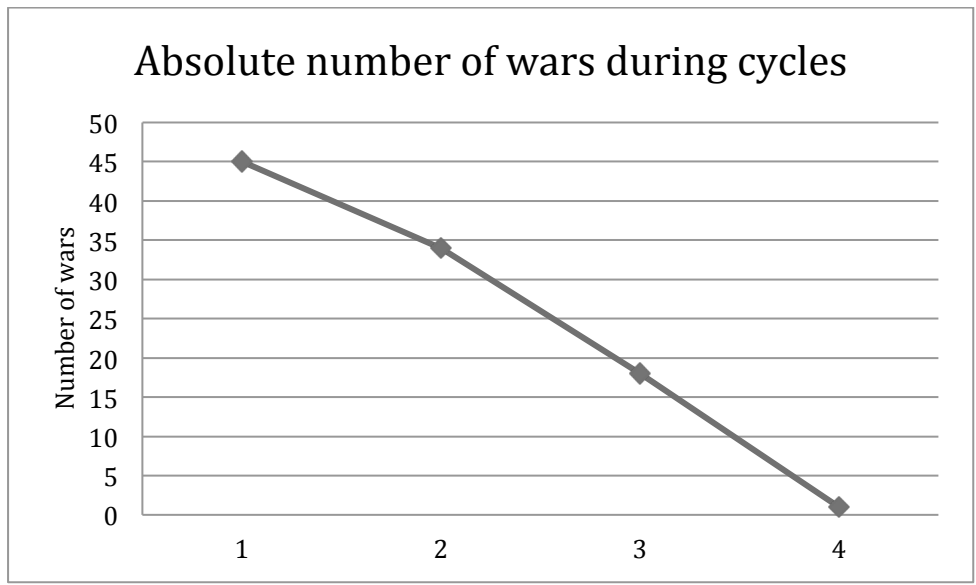

*Figure 6: This figure shows the linear decrease of the resilience of the International System.*

| | **Regularities in dynamics and development** | | | |
|---|---|---|---|---|
| | **Characteristic** | **Type of regularity** | **Mathematical formula** | **Fit (R2)** |
| 1 | Life spans of successive oscillations (singularity) | Exponential decrease | Life-span (t) = 19.6e^(0.936 t) | 0.99 |
| 2 | Stability (based on war frequencies of successive cycles) | Linear increase | War Frequency (t) = 0.465–1.02t | 0.98 |
| 3 | Stability (based on Great Power status dynamics during successive cycles) | Linear increase | Status Dynamics (t) = 10.5–2.6t | 0.99 |
| 4 | Resilience of successive cycles | Linear decrease | Resilience (t) = 61.5–14.8t | 0.99 |
| 5 | Intensity / year of successive systemic wars | Exponential increase | Intensity / year (t) = 5.68 e^(2.61t) | 1.00 |
| 6 | Absolute intensity of successive systemic wars | Exponential increase | Abs. Intensity (t) = 8664.47 e^ (0.62t) | 0.97 |
| 7 | Life-span of successive systemic wars | Exponential decrease | LS(t) = 58.1 e^(-0.61t) | 0.96 |
| 8 | Life-span of successive cycles | Exponential decrease | LS(t) = 194.1 e^(-0.295t) | 0.92 |

*Table 4: An overview of regularities*